\documentclass{article}
\usepackage[final]{colm2026_conference}

\usepackage[T1]{fontenc}
\usepackage{microtype}
\usepackage{hyperref}
\usepackage{url}
\usepackage{booktabs}
\usepackage{graphicx}
\usepackage{wrapfig}
\usepackage{caption}
\usepackage{subcaption}
\usepackage{enumitem}
\usepackage[section]{placeins}

\usepackage{lineno}
\usepackage{listings}
\usepackage{tcolorbox}
\definecolor{jsblue}{rgb}{0.0, 0.0, 0.7}
\definecolor{jsgreen}{rgb}{0.13, 0.55, 0.13}
\definecolor{jsbrown}{rgb}{0.6, 0.2, 0.0}
\lstdefinelanguage{JavaScript}{
  keywords={const, let, var, if, else, for, while, do, return, function, class, new, this, typeof, instanceof, in, of, true, false, null, undefined, async, await, continue, break, throw, try, catch, finally, import, export, default, from, using},
  sensitive=true,
  comment=[l]{//},
  morecomment=[s]{/*}{*/},
  morestring=[b]",
  morestring=[b]',
}
\definecolor{darkblue}{rgb}{0, 0, 0.5}
\hypersetup{colorlinks=true, citecolor=darkblue, linkcolor=darkblue, urlcolor=darkblue}

\usepackage{amsmath,amsfonts,bm}

\def\eqref#1{equation~\ref{#1}}

\def\1{\bm{1}}

\DeclareMathAlphabet{\mathsfit}{\encodingdefault}{\sfdefault}{m}{sl}
\SetMathAlphabet{\mathsfit}{bold}{\encodingdefault}{\sfdefault}{bx}{n}

\title{Composer 2 Technical Report}

\author{\vspace{-0.7cm} \\  Cursor Research Team}

\begin{document}

\ifcolmsubmission
\linenumbers
\fi

\maketitle
\vspace*{-1cm}

\section{Introduction}
\label{sec:intro}

Composer 2 is a specialized model designed for agentic software engineering. The model demonstrates strong long-term planning and coding intelligence while maintaining the ability to efficiently solve problems for interactive use. 
The model scores strongly on CursorBench, our benchmark of real-world software engineering (Figure~\ref{fig:teaser}), while also scoring at frontier levels on public software engineering benchmarks such as SWE-bench Multilingual~\citep{jimenez2024swebench} and Terminal-Bench~\citep{merrill2026terminal}.

The model is trained in two phases: first, continued pretraining to improve the model's knowledge and latent coding ability, followed by large-scale reinforcement learning to improve end-to-end coding performance through stronger reasoning, accurate multi-step execution, and coherence on long-horizon realistic coding problems.

A core tenet of Composer training is to emulate real-world user challenges as closely as possible to minimize train-test mismatch. We develop infrastructure to support training in the same Cursor harness that is used by the deployed model, with equivalent tools and structure, and use environments that match real problems closely. To measure the ability of the model on increasingly difficult tasks, we introduce a benchmark derived from real software engineering problems in large codebases including our own.

Composer 2 is a frontier-level coding model and demonstrates a process for training strong domain-specialized models. On our CursorBench evaluations the model achieves a major improvement in accuracy compared to previous Composer models (61.3). On public benchmarks the model scores 61.7 on Terminal-Bench and 73.7 on SWE-bench Multilingual in our harness, comparable to state-of-the-art systems.

\begin{figure}[]
\centering
\includegraphics[width=1.0\linewidth]{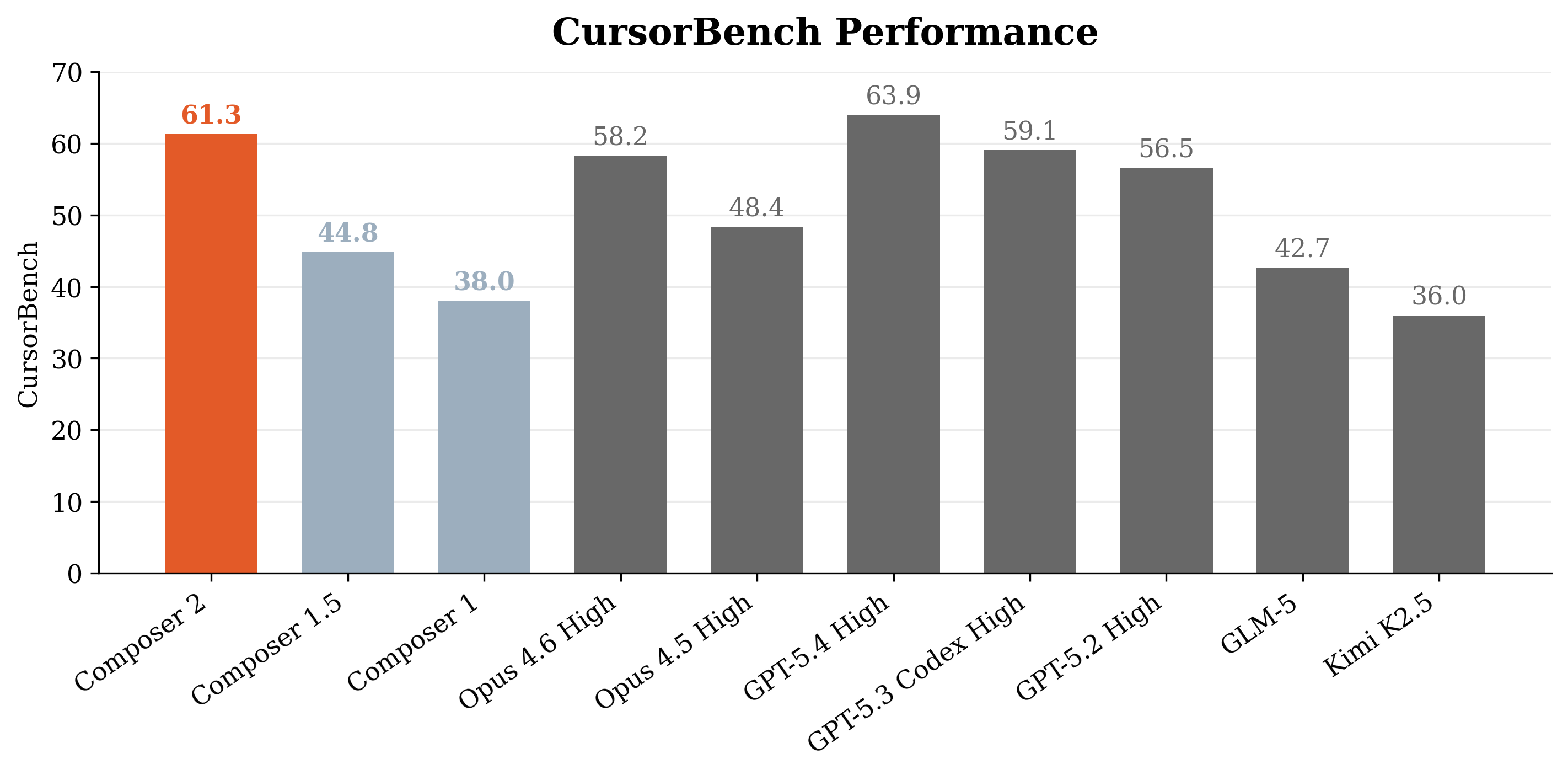}
\caption{\textbf{Composer 2 improves greatly from previous Composer models, achieving performance competitive with state-of-the-art models.} By specializing entirely on coding ability, Composer attains such performance while being lower cost to serve than state-of-the-art model API pricing. See Section \ref{sec:evals} for detailed evaluations.}
\label{fig:teaser}
\end{figure}

\section{Background and Related Work}
\label{sec:background}

Generating code has been a standout application of large language models~\cite{feng2020codebert,clement2020pymt5,chen2021codex,li2022competition}. Code provides a rich source of challenging training data that has supplemented language data in most large models \cite{fried2023incoder,li2023starcoder, lozhkov2024starcoder2,roziere2023codellama,guo2024deepseekcoder,deepseekcoder_v2, allal2023santacoder,nijkamp2023codegen,hui2024qwen2,wang2021codet5,wang2023codet5+,team2024codegemma,mishra2024granite}. Early applications of code generation typically focused on autocomplete applications. Subsequently, instruction tuning turned models into coding assistants \cite{luo2023wizardcoder,wei2023magicoder,zhuo2025parameter,muennighoff2023octopack} capable of responding to user requests. In the last year, software engineering \textit{agents} have achieved widespread adoption, pushing models beyond chat to autonomously navigate repositories and solve complex engineering tasks \cite{yang2024sweagent,yang2025swesmith,wang2024opendevin,qian2024chatdev,hong2023metagpt}.

Software engineering agents aim to autonomously act to solve a given task prompt. Given an environment, i.e., a codebase and an isolated container for code execution, along with a prompt $x$ giving the agent its task, an agent produces a rollout consisting of a series of actions $a_1, \ldots, a_T$, each of which makes one or more tool calls and yields responses $y_1, \ldots, y_T$. Tool calls may modify the underlying environment, and the result of a rollout is the final state of this environment. Each action $a_i$ is selected by sampling from a language model policy $\pi_\theta(a_i \mid x, a_1, y_1, \ldots,a_{i-1}, y_{i-1})$, after which a reward is given based on the code's correctness, succinctness, and conformance to software engineering principles. In contrast to more constrained settings like competitive programming, a strong software engineering agent must perform non-trivial exploration, write its own tests, and construct the minimal changes necessary to solve the task prompt.

Composer 2 has access to a small set of general tools that allow it to read and edit files, run shell commands, search the codebase using grep or semantic search,  and search the web. Its prompt includes a system message, the tool call format specification, recent file information, past user messages, and the current task. The most common end result of this process is a set of changes to files in the codebase environment, although there are many other common use cases, such as answering questions, writing plans, resolving version control issues, or monitoring long-running jobs. 

Our main research thrust for Composer 2 investigates how scaling model training can reliably improve performance on real-world coding. 
We target this through two distinct training phases: continued pretraining (Section~\ref{sec:pretraining}), and asynchronous reinforcement learning (Section~\ref{sec:rl}). To measure progress, we construct a suite of challenging benchmarks (Section~\ref{sec:evals}).

\section{Continued Pretraining}
\label{sec:pretraining}

The continued pretraining stage aims to improve the language model's base knowledge, specifically in the domain of coding. Such continued pretraining has long been demonstrated to drastically improve downstream performance~\cite{gururangan2020dontstop, howard2018ulmfit}. Taking this a step further, recent models use a staged training approach, progressively filtering towards higher quality data \cite{hoffmann2022chinchilla, touvron2023llama, ye2025datamix}. While we start with base models naturally trained with large amounts of code data, we find that additional supervised learning reliably improves knowledge benchmarks and leads to improved coding performance of the final coding agent. 

We used internal evaluations and inference performance considerations to select a base model. Our evaluations measure internal codebase perplexity, coding knowledge, and state tracking. For more details, see Appendix~\ref{sec:basemodelselection}. These evaluations led us to select Kimi K2.5~\cite{kimik2_5}, a 1.04T parameter / 32B active parameter Mixture-of-Experts model as our base model for Composer 2. 

\subsection{Training}
\label{sec:pretraining_training}

We extend Kimi K2.5 with a continued pretraining stage on a large code-dominated data mix. The purpose of this stage is to provide a base model for the subsequent agentic RL training by specializing the model on coding knowledge and capabilities. We divide this stage into three phases. We spend the bulk of compute at 32k token sequence length, followed by a shorter long-context extension phase to 256k sequence length, and finally a short SFT phase on targeted coding tasks. Training was performed in MXFP8 on NVIDIA B300s using the AdamW optimizer. See Section~\ref{sec:infra_training} for more training details. During training, we measure the evaluation loss on our internal codebase. We see that the loss decreases log-linearly over the course of the training run.

Continued pretraining ultimately serves to improve downstream RL performance, and the connection between the two stages is an area of active research. We study the relationship between codebase perplexity and RL performance by applying our continued pretraining recipe to Qwen3-Coder-30B-A3B~\cite{qwen3coder}. Continued pretraining is performed at three logarithmically spaced compute levels: small, medium, and large. Each of these checkpoints then undergoes SFT on a small dataset, followed by an identical RL run. Figure~\ref{fig:pretraining_curves} (left) shows the relationship between the final loss after SFT and the RL reward after a fixed number of steps, demonstrating that cross-entropy loss is indeed predictive of downstream RL performance.

\begin{figure}[t!]
\centering
\includegraphics[width=\linewidth]{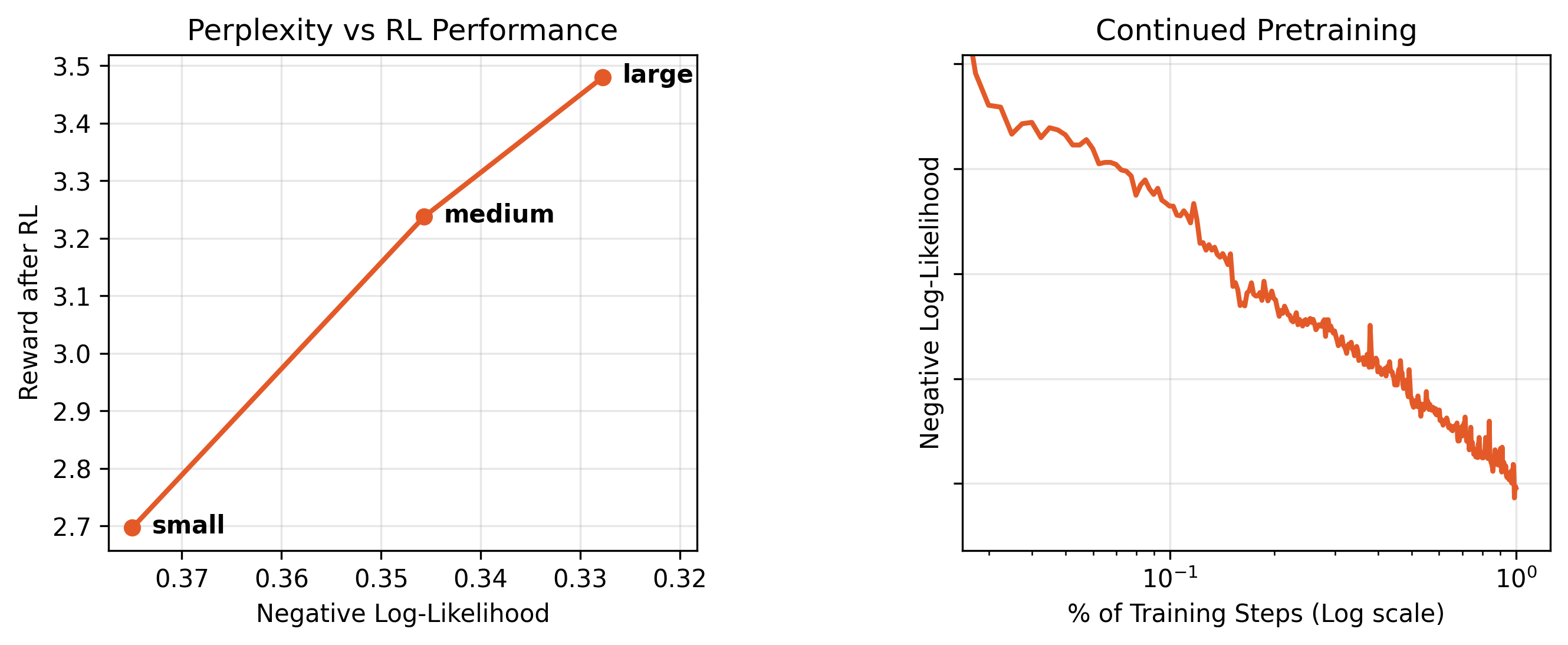}
\caption{\textbf{Continued pretraining translates to downstream RL performance.} Left: We study this relationship on a smaller Qwen model, examining checkpoints trained on a varying number of tokens. Right: The model undergoes a steady decrease in training perplexity.}
\label{fig:pretraining_curves}
\end{figure}

\paragraph{Multi-Token Prediction} To serve the model faster in production, we train additional Multi-Token Prediction (MTP) layers \cite{gloeckle2024mtp, deepseekv3} to use with speculative decoding. We initialize the MTP layers from scratch and train them on the same data mix. To speed up convergence, we train the MTP layers with self-distillation, teaching the model to predict the exact logit distribution of the main LM head at each position. To ensure that this process generalizes, the MTP layers are trained atop a checkpoint cut from the middle of the continued pretraining run. During the final two phases (long-context and SFT), the MTP layers are included and trained jointly with the rest of the model.

\section{Reinforcement Learning}
\label{sec:rl}

\begin{wrapfigure}[13]{r}{0.4\textwidth}
    \centering
    \vspace{-1cm}
\includegraphics[width=1.0\linewidth]{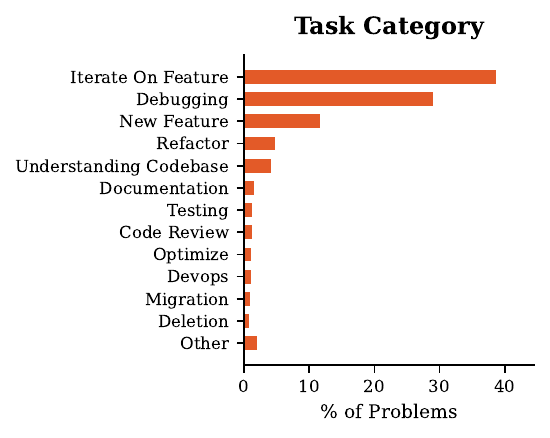}
\caption{RL training tasks.}
\label{fig:task_distribution}

\end{wrapfigure}

Composer 2 is trained by reinforcement learning on a large set of coding tasks. 
These tasks are run in environments that emulate real Cursor sessions as closely as possible (see Section~\ref{sec:infra_rl} for infrastructure details). 
At a high level, RL training consists of sampling a problem, simulating a group of rollouts from the agent with different solutions, and then updating the model weights based on solution quality.

We create a problem distribution that reflects the most common use cases. Figure~\ref{fig:task_distribution} shows the breakdown in terms of task category. Notably, our training distribution captures many aspects of software engineering absent from popular AI coding benchmarks. In later stages of training, we use simple heuristics---such as number of turns and thinking tokens of rollouts---to upsample increasingly harder data points.

\subsection{Asynchronous RL Training}
\label{sec:rl_training}

Our reinforcement learning pipeline is built around learning from large-scale policy gradients while maintaining stability.
We use a policy gradient algorithm with multiple samples per prompt \cite{shao2024grpo, ahmadian2024rloo} and a fixed group size.
We operate in the single-epoch regime, i.e., the same prompt is never trained on twice.
We utilize Adam as our underlying optimizer and update the full parameter set. RL training operates in a highly asynchronous regime with independent training and rollout generation workers (see Section~\ref{sec:infra_rl} for details).

A number of policy gradient variants have been proposed in prior literature \cite{yu2025dapo, zheng2025gspo, minimaxm1, liu2025drgrpo}. As in Dr. GRPO~\cite{liu2025drgrpo}, we found that it is crucial to minimize the bias in the gradients that can arise from transforming the underlying advantage. Following this work, we remove the length standardization term from GRPO as it introduces a length bias. We do not normalize group advantages by their standard deviation, as it results in the degenerate case where small behavioral differences get massively upweighted within a group where every rollout achieves equal correctness.

\citet{yu2025dapo} proposed to mask out rollouts that exceed the maximum sequence length. Some subsequent works employed this masking~\cite{liu2025acereason, golubev2025longcontext}, while other works found it to yield mixed results. For instance,~\citet{liu2025drgrpo} found that masking overlong rollouts shows limited effectiveness on long-tail reasoning tasks but increases the accuracy and clarity of responses in medium and short-length reasoning tasks, and~\citet{du2025ulorl} found that overlong masking caused output length to grow too quickly. We did not see benefits with overlong masking at small scale and opted not to mask rollouts that exceed the maximum sequence length. Our self-summary system (discussed below) also limits the occurrence of these cases in practice. 

Since agent rollouts can be very long, especially when aiming for long-horizon coherency, it is important that our system maintains stability in the highly asynchronous regime. Our main strategy is to minimize how off-policy the samples become. On the infrastructure side, this divergence is reduced via fast weight synchronization and in-flight weight updates, similar to PipelineRL~\cite{pipelinerl}. Inference workers are capable of updating weights mid-rollout, which means later tokens in a rollout are likely less off-policy. To reduce further divergence between the sampling and training policy, we replay MoE routing~\cite{ma2025routerreplay}. We discuss the implementation of our asynchronous RL pipeline in Section~\ref{sec:infra_rl}.

\begin{wrapfigure}{r}{0.4\textwidth}
    \centering
    \includegraphics[width=\linewidth]{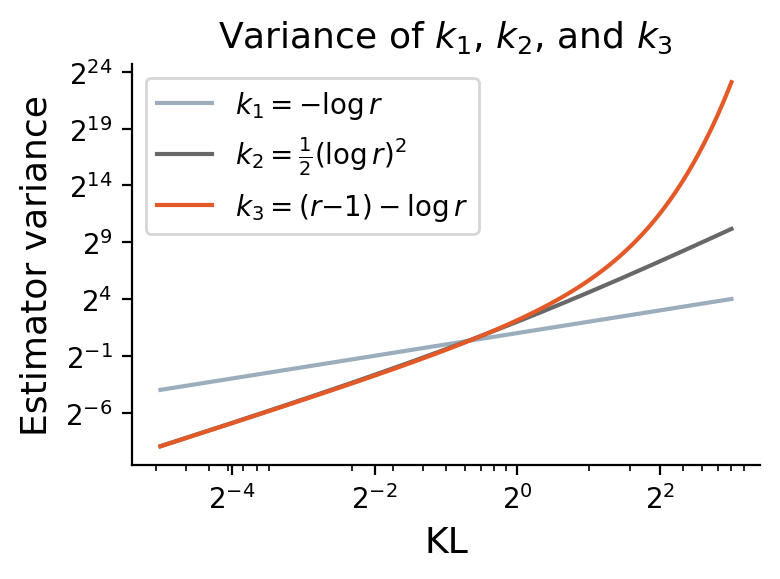}
    \caption{Comparison of estimators of $\mathrm{KL}(p \,\|\, q)$ for two synthetic Gaussian distributions with unit variance and different means.}
    \label{fig:kl-figure}
\end{wrapfigure}

Similar to prior work~\cite{shao2024grpo, kimik1_5}, we use a Kullback--Leibler divergence for regularization, $\mathrm{KL}(q \,\|\, p)
= \mathbb{E}_{x \sim q}\!\left[-\log r(x)\right]$, $r(x) = p(x)/q(x).$
Many open-source implementations of RL estimate KL with the estimator $k_3 = (r-1) - \log r$, defined in~\citet{schulman2020kl}. The $k_3$ estimator is an unbiased estimator of KL and reduces variance when $p$ and $q$ are close. However, Amini et al. shows in~\cite[Figure 1]{amini2025better} that the variance increases drastically as $p$ and $q$ diverge. See Figure~\ref{fig:kl-figure}: for large KL values, the variance of the estimate is extremely large. (The $k_2$ estimator does not suffer from variance blow-up, but is biased.) Therefore, we use the standard estimator $k_1 = -\log r$ instead.

\begin{figure}[t!]
\centering
\includegraphics[width=1.0\linewidth]{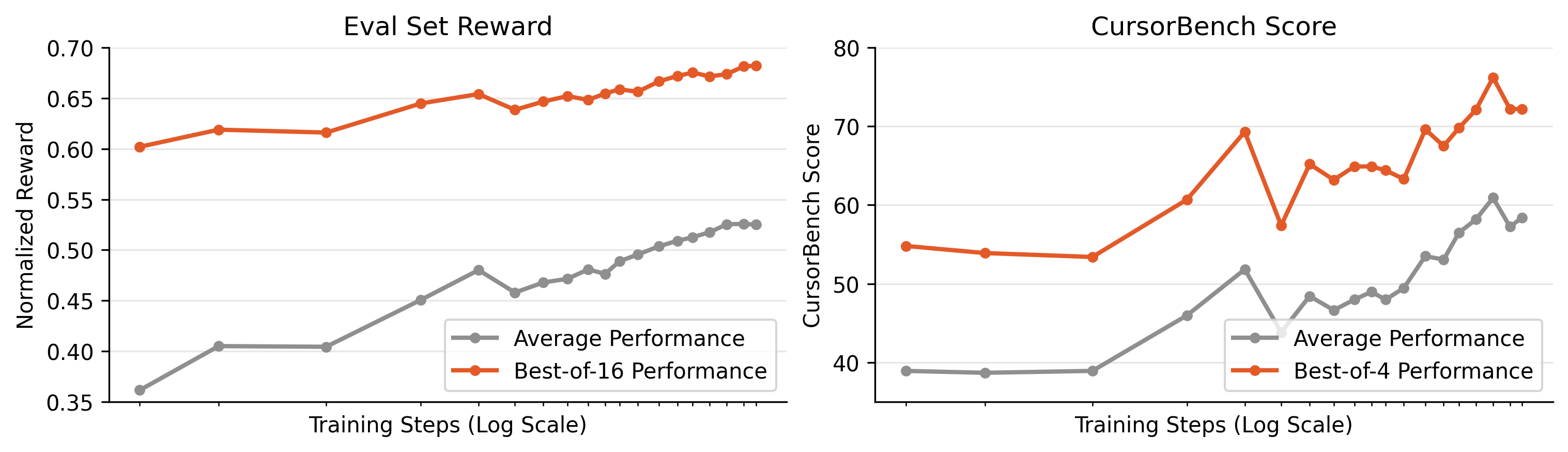}
\caption{\textbf{Both average and best-of-K performance increase over the RL training period.} The above curves are reported on a held-out evaluation set, along with CursorBench tasks. Performance steadily improves throughout RL training. Importantly, we do not observe a tradeoff between average performance and best-of-K performance.}
\label{fig:rl_bon}
\end{figure}

A growing body of recent literature has argued that RL on LLMs often improves average performance primarily by concentrating probability mass on already-known successful trajectories, sometimes at the cost of policy entropy and output diversity~\cite{yue2025doesrlreally, liang2025beyondpass1, chen2025passktraining, wen2025rlvrimplicitly, tajwar2026maxrl}. Under this view, improvements at best-of-K may be limited because the model becomes better at selecting one high-confidence solution rather than expanding the set of reachable correct solutions. Against this backdrop, our results are notable: rather than observing a trade-off in which average reward rises while best-of-K remains flat, we find that our training improves both statistics  as shown in Figure~\ref{fig:rl_bon}. This suggests that, in our setting, RL is not merely reweighting a fixed pool of reasoning paths, but is also improving the model’s effective coverage of correct solutions under repeated sampling.

\paragraph{Self-Summarization}

To enable Composer 2 to work across long horizons, we use the self-summarization technique introduced in Composer 1.5~\cite{cursor_selfsummary}.  Each training rollout can involve multiple generations chained together by summaries, rather than a single prompt--response pair. We use the final reward for all tokens produced by the model in the chain. This upweights both the agent responses in good trajectories and also the self-summarizations that made them work. At the same time, poor summaries that lose critical information are downweighted. As Composer trains, it learns to use self-summaries to process more information, even with a limited context window. For hard examples, it often self-summarizes multiple times. In our experiments, we find that self-summary consistently reduces the error compared to using separate prompt-based compaction, while using significantly fewer tokens and reusing the KV cache.

\subsection{Agent Behavior}

While the primary goal of RL training is to improve model intelligence, we also aim to produce a model that provides a good developer experience. 
This is affected by the communication style of the model as well as the time and resources it takes to answer a question. 

\begin{wrapfigure}{r}{0.4\textwidth}
    \centering
    \includegraphics[width=\linewidth]{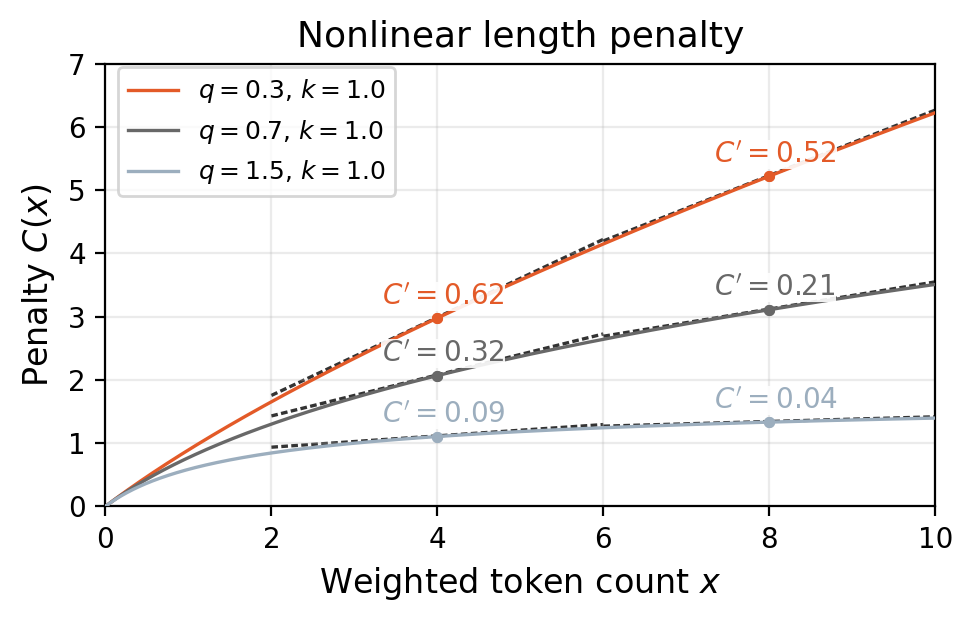}
    \caption{
    Nonlinear penalties push the model to be quick on easy tasks and think more on hard tasks.
    }
    \label{fig:nonlinear}
\end{wrapfigure}

For behavior and communication, we apply an array of auxiliary rewards to ensure the model provides a good experience. These include rewards for coding style, communication, and product-specific penalties for poor tool calls, such as creating to-do list items and then leaving them unfinished. During  RL training, we monitor the model for emergent behaviors and occasionally introduce additional behavior rewards as needed. For example, we observed that the model would start to leave long chains-of-thought in comments or collapse to using the terminal tool only.

To incentivize the model to produce solutions quickly on easy requests while allowing it to think longer on hard requests, we add a concave down and increasing nonlinear length penalty to the reward:
\begin{equation*}
C_{\text{length} \{k, q\}}(x) = \frac{(1+kx)^{1-q}-1}{k(1-q)},
\label{eq:length_penalty}
\end{equation*}
where $k$ and $q$ are hyperparameters which define the curvature of the penalty, and the input $x$ is a weighted combination of thinking tokens, tool calling tokens, tool output tokens, final message tokens, number of tool calls, and number of turns of a rollout.
The nonlinearity reflects that on easy tasks, achievable with only a few tool calls, every additional bit of effort is felt more acutely than in long-horizon tasks, where the agent might iterate for hundreds of tool calls. See Figure~\ref{fig:nonlinear} for some examples of the nonlinear curves produced by this equation.
We find that utilizing such length penalties enables the model to learn particularly efficient behaviors, e.g., making multiple tool calls in parallel.

\section{Real-World Evaluation with CursorBench}
\label{sec:evals}
The application of coding agents has evolved rapidly over the past year, expanding from simple, tightly-scoped edits to complex debugging, large-scale refactoring, and feature development.
At Cursor, we have observed that performance on public evaluation benchmarks often correlates only loosely with the real-world utility of these models.
We attribute this misalignment to four primary factors:

\begin{itemize}[leftmargin=*]
\item \textit{Domain Mismatch:} 
As the capabilities of coding agents expand, static benchmarks often fail to capture the full spectrum of developer workflows.
For instance, SWE-bench and its variants predominantly focus on isolated bug-fixing.
Terminal-Bench covers a wider range of task types, but many of its tasks (e.g., computing chess moves) are abstract puzzles rather than typical software engineering operations.

\item \textit{Prompt Over-specification:} 
Public benchmarks are typically highly specified, assuming a narrow set of correct solutions.
In contrast, real developer requests are often underspecified and admit multiple valid architectural approaches.
Consequently, public benchmarks either penalize correct alternative solutions or rely on unnaturally explicit prompts that bypass the challenge of interpreting ambiguous intent.

\item \textit{Data Contamination and Overfitting:} 
Because public benchmarks are constructed from historical scrapes of open-source repositories, they are frequently leaked into model training mixtures, artificially inflating scores.
Recently, OpenAI suspended reporting SWE-bench Verified results after finding evidence that frontier models could generate gold patches from memory~\cite{openaiSWEbenchVerified}.
Beyond contamination, the fixed and narrow nature of these benchmarks can compress performance differences: for instance, Haiku 4.5 achieves 73.3\% on SWE-bench Verified, very close to GPT-5's 74.9\%, misaligning with accuracy on broader and more diverse task distributions like Terminal-Bench.

\item \textit{Narrow Evaluation Scope:} 
Existing coding evaluations predominantly measure functional correctness.
In practice, developers also heavily weigh code quality, readability, latency, cost, and the quality of the agent's interactive behavior throughout a session.
\end{itemize}

\begin{figure}[t!]
\centering
\begin{subfigure}[t]{0.48\linewidth}
\centering
\includegraphics[width=\linewidth]{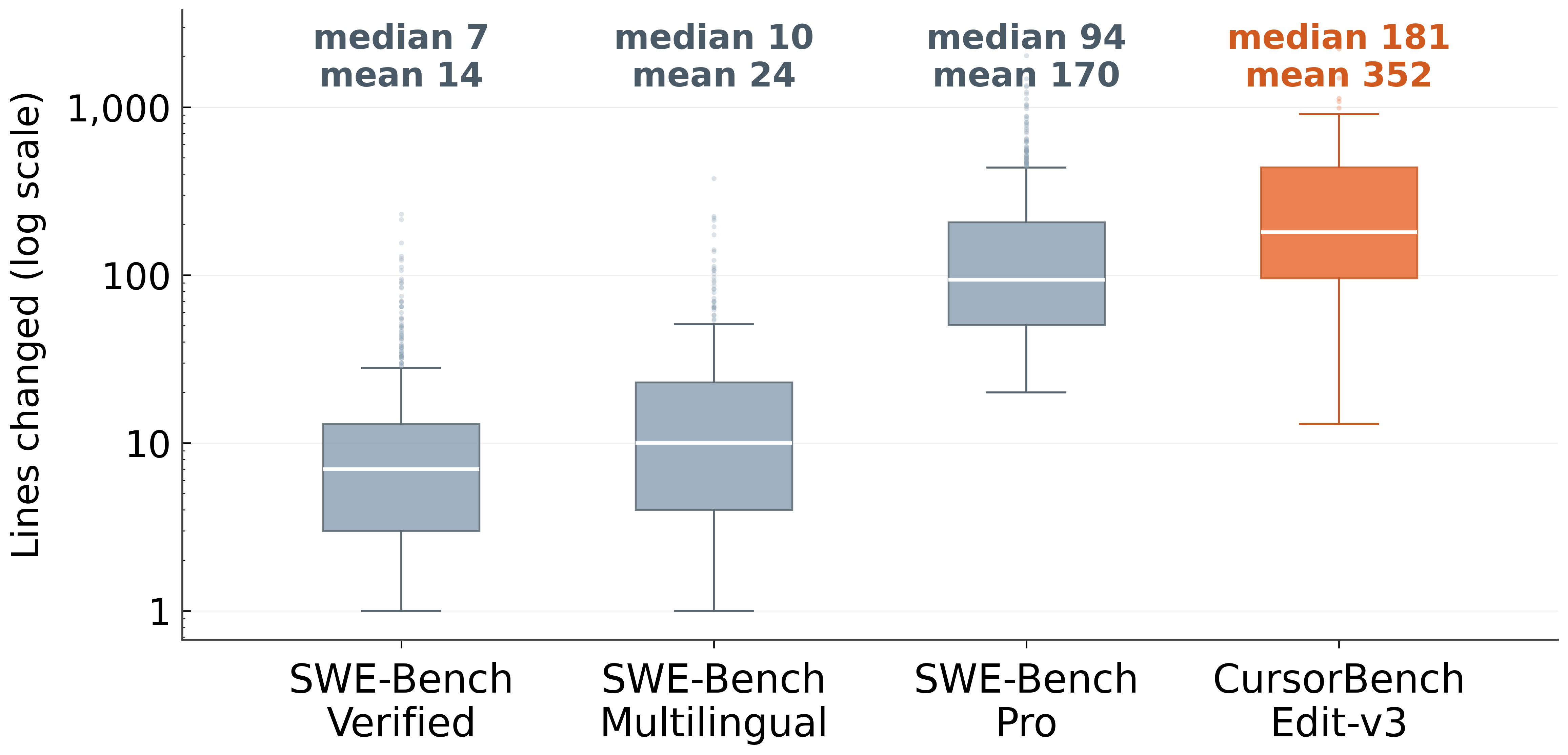}
\caption{Lines changed in reference diff.}
\label{fig:lines_changed}
\end{subfigure}
\hfill
\begin{subfigure}[t]{0.48\linewidth}
\centering
\includegraphics[width=\linewidth]{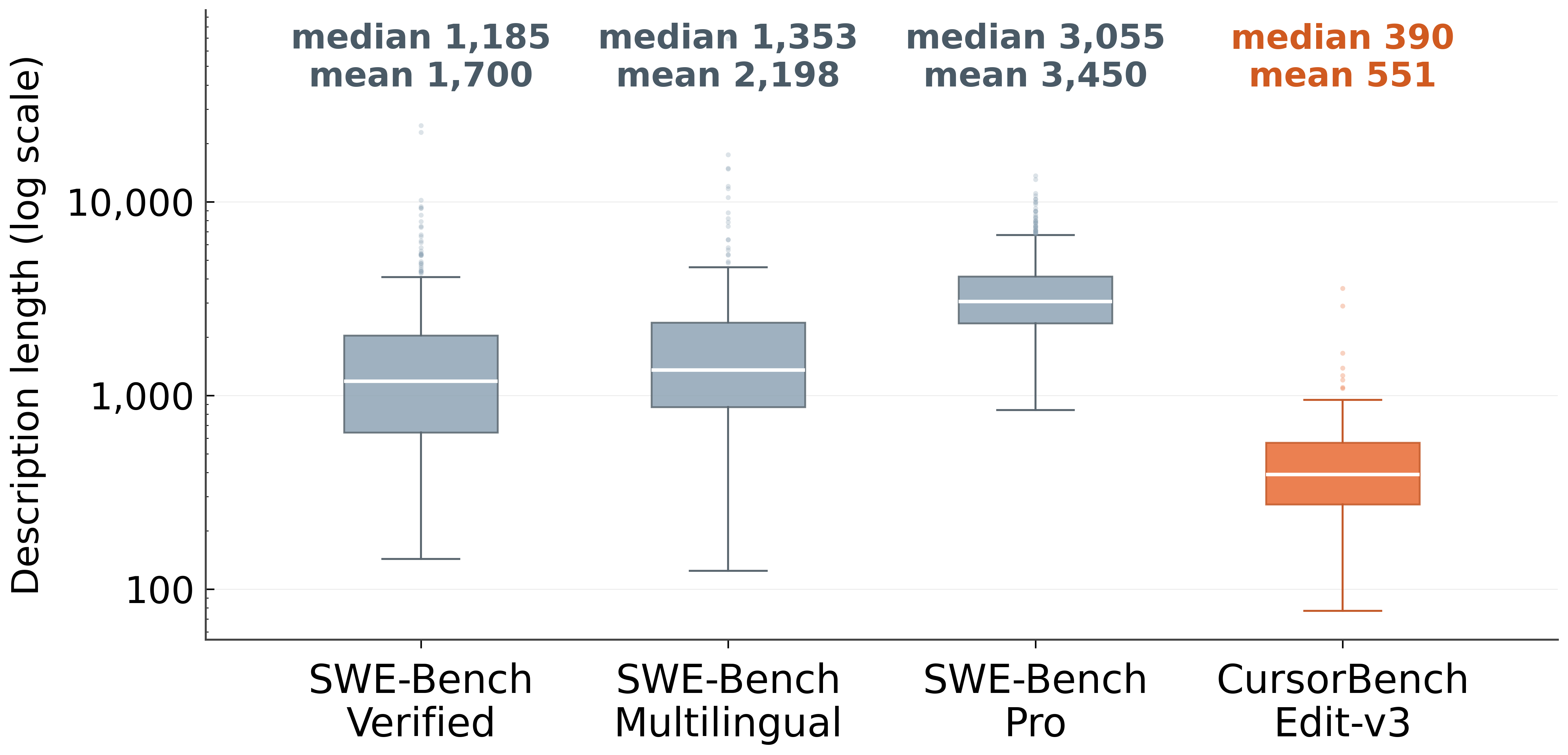}
\caption{Problem description length.}
\label{fig:desc_length}
\end{subfigure}
\caption{\textbf{Compared to public benchmarks, CursorBench tasks have less-specified task prompts, and require an order of magnitude more code changes.} We find this better represents the complexity and ambiguity of real-world software engineering requests.}
\label{fig:cursorbench_comparison}
\end{figure}

To address these limitations, we introduce CursorBench, an internal evaluation suite comprising tasks drawn from actual coding sessions of our engineering team.
Because these tasks originate from real agent sessions rather than curated public repositories, CursorBench better reflects the true distribution of software engineering tasks while completely avoiding train-set contamination.
Furthermore, rather than relying solely on functional correctness, we evaluate models using specific metrics targeting code quality, execution efficiency, and interactive agent behavior in realistic settings.

Figure~\ref{fig:cursorbench_comparison} highlights the structural differences between CursorBench and public evaluation sets.
CursorBench tasks necessitate substantially more extensive code modifications, with a median of 181 lines changed compared to just 7--10 lines for SWE-bench Verified and Multilingual (Figure~\ref{fig:lines_changed}).
At the same time, CursorBench prompts are also more underspecified, featuring a median description length of only 390 characters versus 1,185--3,055 characters for public benchmarks (Figure~\ref{fig:desc_length}).
This combination of broad execution scope and high intent ambiguity accurately reflects the intrinsic difficulty of real-world software engineering, where developers must frequently synthesize context from production logs, sparse user bug reports, and large existing codebases to derive a solution.
Figures~\ref{fig:cursorbench_example} and \ref{fig:cursorbench_example_streaming} show representative examples: one requires diagnosing a build-tool transpilation bug in a retry loop from a terse bug report and observability logs, while the other requires designing a tuned heuristic detector over hundreds of chat responses to quantify a subtle streaming regression and discover its hidden invariants.

\begin{figure}[t!]
\centering
\begin{lstlisting}[language=JavaScript, basicstyle=\ttfamily\scriptsize, escapechar=|]
// executeScoringRollout.ts - linked code snippet from the problem statement

for (let attempt = 1; attempt <= MAX_RETRIES; attempt++) {
  try {
    const request = new ScoringRequest(...);
    const { ctx: Ctx, startSpan: taskSpan } = ctx.span("scoring");
    |\colorbox{red!15}{\texttt{using \_taskSpan = taskSpan.start();}}|
    const result = await executeScoring(...);

    let rawOutput = "";
    if (result.response) {
      rawOutput = result.response.join("\n");
    }
    const parsed = parseOutput(rawOutput);
    if (parsed.parseError) {
      lastError = parsed.parseError;
      ctx.warn({ error: lastError }, "Error, will retry");
      if (attempt < MAX_RETRIES) { continue; }
    }
    // ...
  } catch (error) { /* ... */ }
}
\end{lstlisting}
\vspace{0.5em}
\begin{tcolorbox}[colback=gray!5, colframe=gray!50, boxrule=0.5pt, left=6pt, right=6pt, top=4pt, bottom=4pt]
\small\textbf{Problem statement:} scoring attempt 2 and attempt 3 succeeded but i get ``failed after 3 attempts. Last error: {[}canceled{]} User aborted request'' error at the end
\newline\newline
@executeScoringRollout.ts (1084-1118)\newline
check if there is some bug in this
\newline\newline
Please see datadog logs at @logs and fix
\end{tcolorbox}
\vspace{-0.5em}
\caption{\textbf{Example CursorBench task} (truncated and obfuscated from our evaluation pipeline). The agent receives a terse bug report and must cross-reference the source code with production observability logs to diagnose the failure. The logs also contain unrelated production service warnings which are a red herring: the true root cause is an esbuild 0.20.2 downleveling bug for \texttt{using}. The transpiled output lowers the highlighted declaration into \texttt{var}-scoped error state that is not reset between retry iterations, causing stale failure state to be re-thrown from the generated \texttt{finally} block even after later attempts succeed.}
\label{fig:cursorbench_example}
\end{figure}

\begin{figure}[t!]
    \centering
    \begin{subfigure}[t]{0.48\linewidth}
    \centering
    \includegraphics[width=\linewidth]{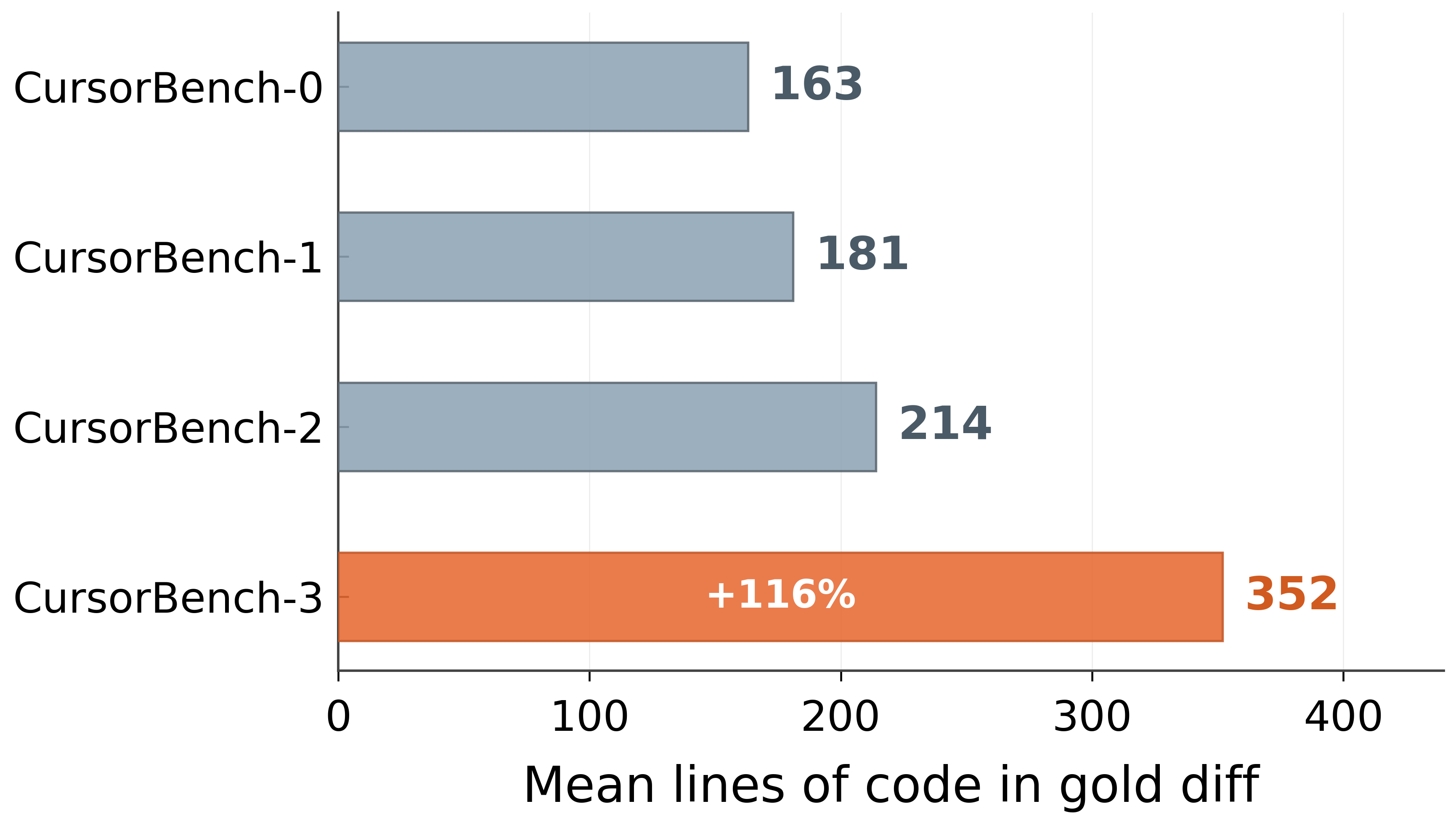}
    \label{fig:iter_lines}
    \end{subfigure}
    \hfill
    \begin{subfigure}[t]{0.48\linewidth}
    \centering
    \includegraphics[width=\linewidth]{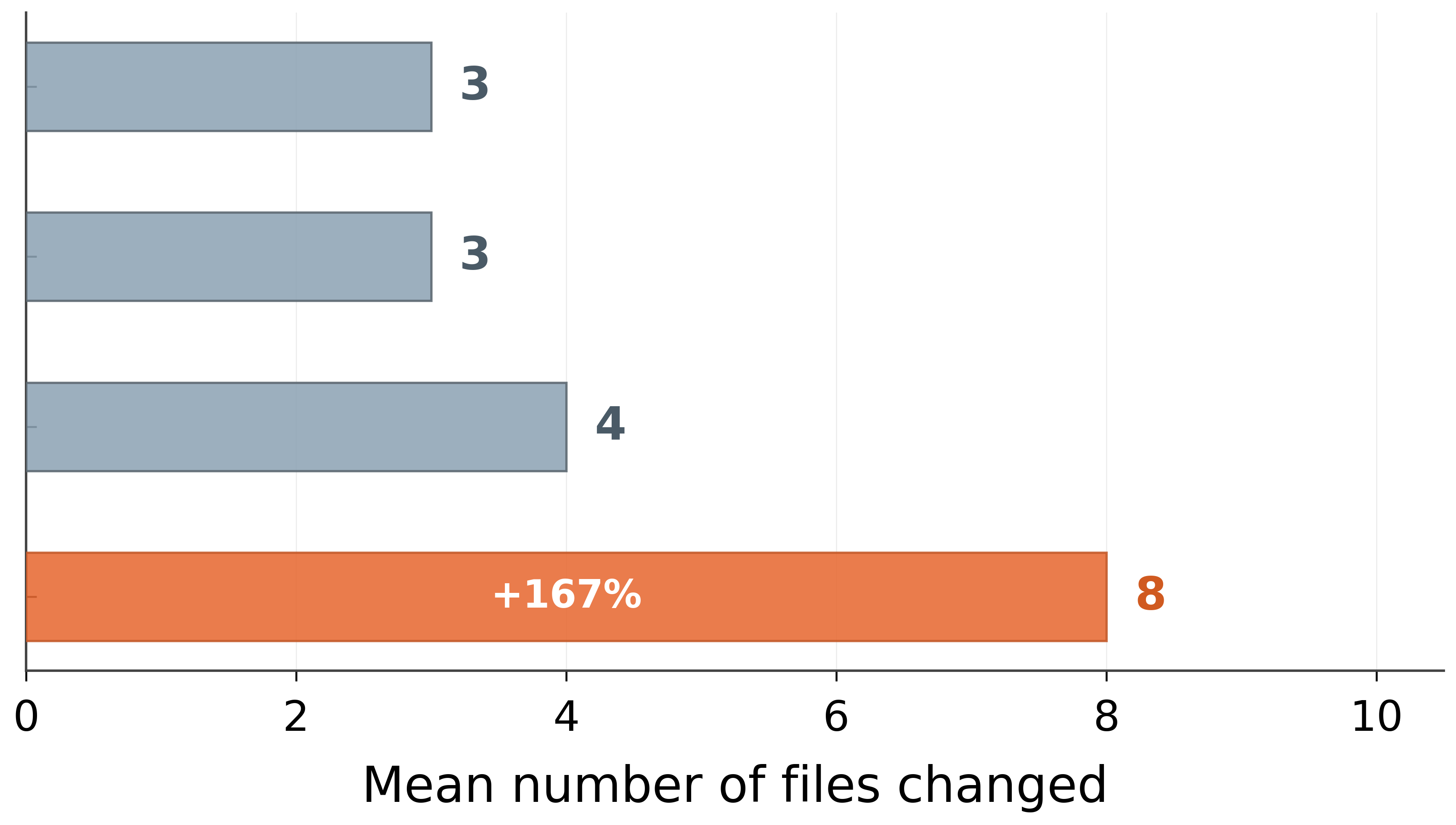}
  
    \label{fig:iter_files}
    \end{subfigure}
    \caption{\textbf{Evolution of CursorBench across iterations.} Each version incorporates more complex requests. CursorBench-3 more than doubles the median task size from the initial version, shown as the relative percent change in the bottom bar.}
    \label{fig:cursorbench_iterations}
    \end{figure}

New CursorBench iterations are continually developed by our team.
As user workflows evolve and agent capabilities improve, we regularly update the evaluation set to remain aligned with how developers actually use the product.
Figure~\ref{fig:cursorbench_iterations} shows how the benchmark has grown in complexity across iterations: compared to earlier versions of CursorBench, tasks from CursorBench-3 involve changing more than twice as many files and lines of code on average.
In addition to increased problem size, the distribution of task types has also shifted, as developers increasingly delegate long-running command execution, experiment monitoring, and data analysis to agents.
This continual refresh ensures that our evaluations remain aligned with the shifting frontier of real-world difficulty and not saturated.

Finally, we complement our primary CursorBench evaluation with a suite of targeted evaluations covering other aspects of coding agent quality and behavior. These include an intent evaluation, which assesses how the model handles ambiguous prompts; an instruction-following evaluation, which measures how well the model follows system prompts, user prompts, rules, and skills; an eager editing evaluation, which tests how the model responds to questions where it should avoid editing code; a code quality evaluation, which judges the quality of both code and comments; and an interruption evaluation, which quantifies how well the model handles mid-rollout interruptions and user feedback. We develop these evaluations by identifying important dimensions of agent behavior, selecting data points that elicit them, and writing rubrics to measure performance.

\section{Infrastructure}
\label{sec:infra}

\subsection{Training Infrastructure}
\label{sec:infra_training}

\paragraph{Parallelism.}
Previous Composer training stacks combined Fully Sharded Data Parallelism (FSDP)~\cite{rajbhandari2020zero, zhao2023pytorch}, Expert Parallelism (EP)~\cite{shazeer2017outrageously, fedus2022switch}, and Tensor Parallelism (TP)~\cite{shoeybi2019megatron}.
In the original MoE design, EP reused the same rank group as TP, so EP was not an independent scaling axis.
This coupling kept the implementation simple, but constrained support for larger MoE configurations and would unnecessarily enable activation sharding in the continued pretraining phase, even when activation memory pressure is modest.

Composer 2 instead uses Context Parallelism (CP)~\cite{liu2024ringattention, jacobs2023deepspeed} as the primary long-context scaling axis. CP requires less communication than TP and improves compute efficiency by preserving full hidden dimensions in various projections; in contrast, TP produces less efficient skinny local matrix multiplications. There are a few tricks we use to implement CP efficiently in the Multi-Head Latent Attention (MLA) architecture. To minimize communication overhead, we compute local KV latent vectors, all-gather the latent vectors across CP ranks, and then compute the KV projections. Although this replicates the projection on all CP ranks, the projection is small and reduces CP communications, allowing us to fully overlap CP communications with the computation of the Q projection. Additionally, while naive CP causes load imbalance during causal attention as later tokens have to attend to more tokens, we use the technique from \citet{liu2024ringattention} to address this: we split the sequence into $2 \times \text{CP}$ chunks, and the $i$-th rank processes chunks $i$ and $2 \times \text{CP} - 1 - i$, resulting in roughly equal work during causal attention for all ranks. Finally, the context parallelism dimension is folded into the FSDP dimension, allowing us to use CP ranks to reduce per-GPU parameter/state memory usage.

Composer 2 also introduces a more flexible expert-parallel design by decoupling EP from TP. This requires using different meshes for sharding dense layers and expert weights. EP is formed from DP and CP capacity, enabling support for larger expert-parallel degrees and making expert-grouped GEMMs more efficient with larger per-rank token batches. We use EP=8, CP=2 for the continued pretraining phase and EP=8, CP=8 for the RL phase. We use DeepEP to implement high-throughput token dispatch/combine~\cite{deepep2025}. DeepEP communication buffers have relatively low overhead, and DeepEP's kernel uses 20 SMs by default, leaving headroom for concurrent compute. We also quantize the tokens to MXFP8 (discussed below) before dispatch for more efficient communication, which does not affect our precision since we already perform our expert computations in MXFP8. We keep the combine at BF16 for increased precision. To maximize compute--communication overlap, tokens are split into microbatches and pipelined across separate communication and compute streams.

Finally, we found that it was critical for different DP ranks to have similar amounts of compute to achieve high utilization. In continued pretraining, DP balance is easily achieved with fixed sequence lengths. In RL, different rollouts of different prompts can result in very different sequence lengths, so before each training step, we run a global sequence packing stage to ensure balanced DP compute load. The packing algorithm takes into account the increased attention costs of longer sequences.

\paragraph{Kernels.}

\begin{figure}[t]
\centering
\includegraphics[width=0.8\linewidth]{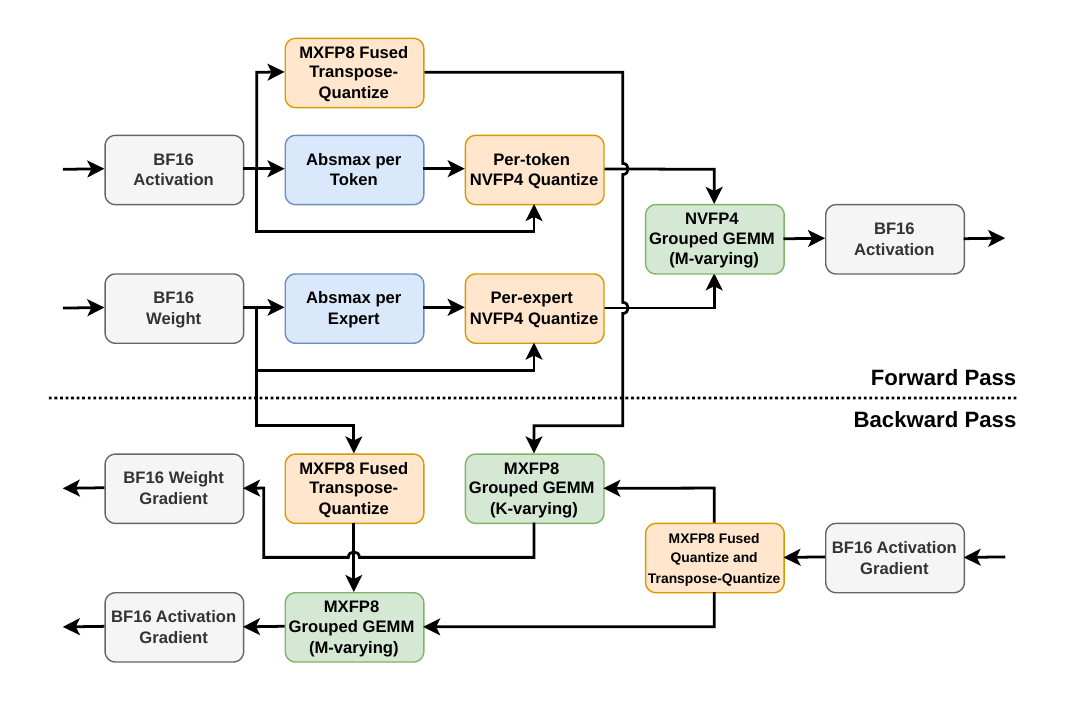} 
\caption{Overview of a single grouped GEMM training flow in our Mixture-of-Experts layer. Each colored block represents a single kernel launch.}
\label{fig:moe-flow}
\end{figure}

Composer 2 training uses in-house kernels written in CUDA, PTX, and ThunderKittens/ParallelKittens~\cite{25-thunderkittens, 25-parallelkittens}. The kernels primarily optimize low-precision training of the mixture-of-experts (MoE) layer. Our training recipe uses both MXFP8~\cite{ocp_mx_spec} and NVFP4~\cite{nvfp4_recipe} precision formats. We exclusively target NVIDIA Blackwell GPUs for block-scaled tensor-core matrix multiplications (i.e., in-hardware dequantization during systolic-array matrix multiplication). Figure~\ref{fig:moe-flow} illustrates a single grouped GEMM training flow within our MoE layer.

For the MoE forward pass, we use a novel variant of NVFP4: values are quantized from BF16 into FP4E2M1 using FP8E4M3 per-block scales (block size = 16) and FP32 \textit{per-token} scales. We found the original NVFP4 format, which uses FP32 per-tensor scales, fragile for two reasons. First, per-tensor scaling makes training batch-variant, collapsing numerical precision and causing the RL training to diverge. Second, inter-token scale values leak future token information into past tokens, resulting in biased gradients. Despite adding latency to the quantization and GEMM epilogue, per-token scaling thus proved to be the more effective scheme.

For the MoE backward pass, we use the standard MXFP8 format with FP8E4M3 values and FP8E8M0 scales per 32-element block. We can do this because of the asymmetry in RL training. On the forward pass, it is necessary that the trainer match the inference for numerical stability. We therefore use trainer NVFP4 in order to support fast inference. The backward pass, however, runs only on the training cluster. This is not a bottleneck on system-wide RL efficiency, so we can afford higher precision to improve training stability.

Finally, the choice of hardware-level math precision mattered considerably. For NVFP4 quantization, we found that using IEEE-compliant floating-point arithmetic (e.g., \texttt{\_\_fdiv\_rn}) is critical; using fast-approximation alternatives causes training to diverge after roughly a hundred RL steps. Conversely, using the fast-approximation path (e.g., \texttt{\_\_fdividef}) for MXFP8 quantization has not caused any divergence since our initial training of Composer 1, so we select it for the best performance.

We actively open-source our kernel implementations and support community efforts to improve the GPU kernel ecosystem. We collaborated closely with Colfax to implement the Flash Attention 4 backward kernel for the QK 192 / V 128 configuration (a.k.a. the "DeepSeek shapes"), which has been merged into the public repository~\cite{flashattention_pr2270}. We also actively support the development of ThunderKittens in collaboration with the Hazy Research group at Stanford ~\cite{25-parallelkittens, sul2025onekernel, sul2025fluffykittens, sul2026thunderkittens2}. Recently, we open-sourced the state-of-the-art BF16, MXFP8, and NVFP4 GEMM implementations into ThunderKittens~\cite{thunderkittens_gemm_kernels}. Finally, we share our knowledge on quantization and MoE kernel implementation through online posts~\cite{cursor_kernels}.

\subsection{RL Infrastructure}
\label{sec:infra_rl}

Our RL infrastructure consists of four decoupled services: training, environments, inference, and evaluations. A decoupled service stack enables larger-scale global training, high availability, and independent scaling and sharding. The production training job for Composer 2 spanned 3 regions for GPU compute and 4 regions for CPU compute.

\paragraph{Training}
We use a fully asynchronous, high-throughput training stack built on Ray ~\cite{moritz2018ray} and PyTorch ~\cite{paszke2019pytorch}. A centralized reconciler performs slot-based sample lifecycle state management, moving samples through a pipeline of distributed executors and implementing scheduling policies that balance sample generation throughput with policy staleness. We design all services within the trainer around the concept of futures, which allow for eager execution of computation when upstream dependencies are ready. We leverage the Ray object store to hold samples that are ready for consumption by train workers, which allows for natural spilling to local NVMe storage when nodes have insufficient CPU memory.

To support large-scale post-training, all components within the trainer are fault-tolerant down to the process or process-group level. We run passive and active health checks on all nodes during training; upon detection of a hardware fault, we mark the node as unhealthy for scheduling but continue training with warm standby nodes. Decoupling training from inference and environment infrastructure naturally makes training more resilient to failures in these services; during the training run, we saw many cases where these services had partial or full outages without failing the training job. To minimize the number of training job restarts, we use a reactive configuration system and support live code updates on a per-process level; when new code is deployed, existing actors are drained of in-flight requests and transparently replaced.

Replaying long-running coding rollouts is expensive. To mitigate expensive failures on job-level faults, we perform policy-aware checkpointing at the rollout level and group level in addition to conventional checkpointing of model weights at the step level. For rollout checkpointing, we rely on memory snapshots of the codebase environment state, so that upon recovery, we can pass the reconstructed codebase environment to verifiers. For group checkpointing, we write sequences with advantages tagged with policy versions to NFS; upon job restart, the scheduler considers these when determining whether to dispatch new work or simply load ready groups.

\paragraph{Environments and Anyrun.}
Stateful codebase environments are a first-class artifact of our post-training stack. Environments are run on top of Anyrun, an internal compute platform built for running untrusted code at scale. This is the same compute platform that powers Cloud Agents and Automations in the Cursor product.

All environment creation requests from the trainer are sent to a global service, which routes the request to an underlying Anyrun cluster. Our training workload is sharded across multiple Anyrun clusters for both instance availability and fault tolerance. Within a cluster, a distributed set of Anyrun managers schedule pods, scale cloud compute provisioned across multiple regions, and perform state reconciliation to manage hundreds of thousands of pods per cluster. Each pod is a dedicated Firecracker VM capable of running a full development environment, including a browser and GUI for computer use. We run pods on a large mixture of machine types and architectures (x86, ARM) to maximize instance availability.

Scheduling throughput is particularly important for the bursty nature of RL workloads. Each Anyrun cluster is capable of scheduling more than 500 pods per second while maintaining desired binpacking requirements. One challenge with a naive packing strategy is that the steady-state resource usage for a pod can be dramatically lower than its peak during startup and can also be bursty due to overcommits. To solve this, we monitor and schedule with awareness of live readings of hardware pressure (CPU, memory, disk) along with more conventional scheduling heuristics.

Anyrun supports forking and snapshotting of full coding environments at both the filesystem and memory level. This unlocks useful capabilities during RL, such as mid-trajectory rollout checkpointing and post-rollout state capture for future introspection. When a pod fork is requested, we attempt to first schedule the fork onto the same node; if not feasible due to space constraints, we live-migrate pod state to a node with capacity.

Egress is carefully controlled in environments to limit any external impact. Any access to the internet from a pod must go through Anygress, an internal service within Anyrun responsible for proxying traffic, enforcing granular request policies, and dropping sensitive headers. To better replicate real-world environments, Anygress operates transparently instead of relying on proxy environment variables by injecting a trusted root CA on pod startup and redirecting pod traffic at the TCP layer.

We train with tools that are representative of the harness in the Cursor client. Each codebase environment starts with a shared tool library that can be invoked over RPC. Some tools like semantic search have external dependencies and are handled outside of the environment. To support the full tool set available in the Cursor client, we maintain a shadow deployment of the Cursor backend that is used both during dataset preparation and rollouts. Sharing the production implementation in this way allows us to scale experiments and training safely while remaining faithful to the harness that Composer 2 will be deployed into.

There are cases where we want tool behavior to differ between training and production settings. Concrete examples include enforcing stricter tool argument checks to encourage more precise model behavior, and removing certain tools to improve model steerability. To achieve this, the set of available tools and the desired behavior of each tool are dynamically determined for each environment.

\paragraph{Inference and Weight Sync.}
We partner with Fireworks AI to run RL inference. Because Kimi K2.5 is a Mixture-of-Experts model, numerical differences can cause different experts to be chosen in the inference engine forward pass and trainer forward pass. If the trainer and inference engine do not agree on expert routing for each token, log-probabilities computed during training may not match the distribution from which tokens were sampled, introducing noise into the policy gradient. To address this, we employ router replay \cite{zheng2025gspo, ma2025routerreplay}: during inference, the engine returns the selected expert indices for every token at every MoE layer, and during the training forward pass the router's expert assignment is overridden to match. The router still computes gating scores so that gradients flow through it. We extend the basic replay scheme by filtering out replayed experts whose gating scores fall below a plausibility threshold derived from the router's own top-$k$ selections, replacing them with the router's candidates; we found that this reduces p99 numerics mismatch between the inference and training forward passes.

Every training step, we synchronize updated weights to the inference engine by uploading to a shared S3 bucket. To minimize transfer size, we use delta compression: each rank caches its previous upload and transmits only the diff against the new weights. Because RL updates are small, even with full-parameter training these diffs compress to a handful of gigabytes for the 1T-parameter model. Uploading is fully sharded across all training ranks, allowing us to saturate the egress bandwidth of the training cluster; similarly, download on the Fireworks side is sharded across inference replicas. Compression, upload, and hotload signaling are fully pipelined in background workers so that training is never blocked. During the Composer 2 training run, we ran inference across geographically distributed clusters in the US and Europe. Each cluster independently downloads and reconstructs weights from the shared delta chain, requiring no direct connectivity to the training cluster, enabling world-scale distributed RL inference over commodity cloud storage.

\paragraph{Online Evaluations.}
To provide faithful evaluations of our model during training, we run a pinned version of the production backend and Cursor client for each evaluation job. This provides high confidence that model behavior during evals is an exact replication of what our end users see, and also allows us to iterate on the Cursor harness and model system prompt using the same infrastructure. For each training step we want to evaluate, we acquire a lease for an evaluation deployment, automatically move GPUs to that deployment, and perform a cross-region weight sync of the evaluation checkpoint from the training cluster where it resides to the inference deployment.

\section{Results}
\label{sec:results}

\subsection{CursorBench}
\label{subsec:cursorbench}
We evaluate our models by running Cursor agents directly within Anyrun (Section~\ref{sec:infra_rl}), the same infrastructure that supports our reinforcement learning pipeline.
For each task in CursorBench, we initialize the codebase environment and initial task prompt, and we run the agent exactly as it would execute in our production environment.

\paragraph{Metrics.} 
We compute accuracy aggregated over all tasks across multiple passes of the evaluation set to reduce variance.
In addition to accuracy, we also measure efficiency metrics like completion tokens, end-to-end latency, and inference cost to ensure the model remains maximally useful for interactive developer workflows.

\begin{figure}[t!]
\centering
\includegraphics[width=\linewidth]{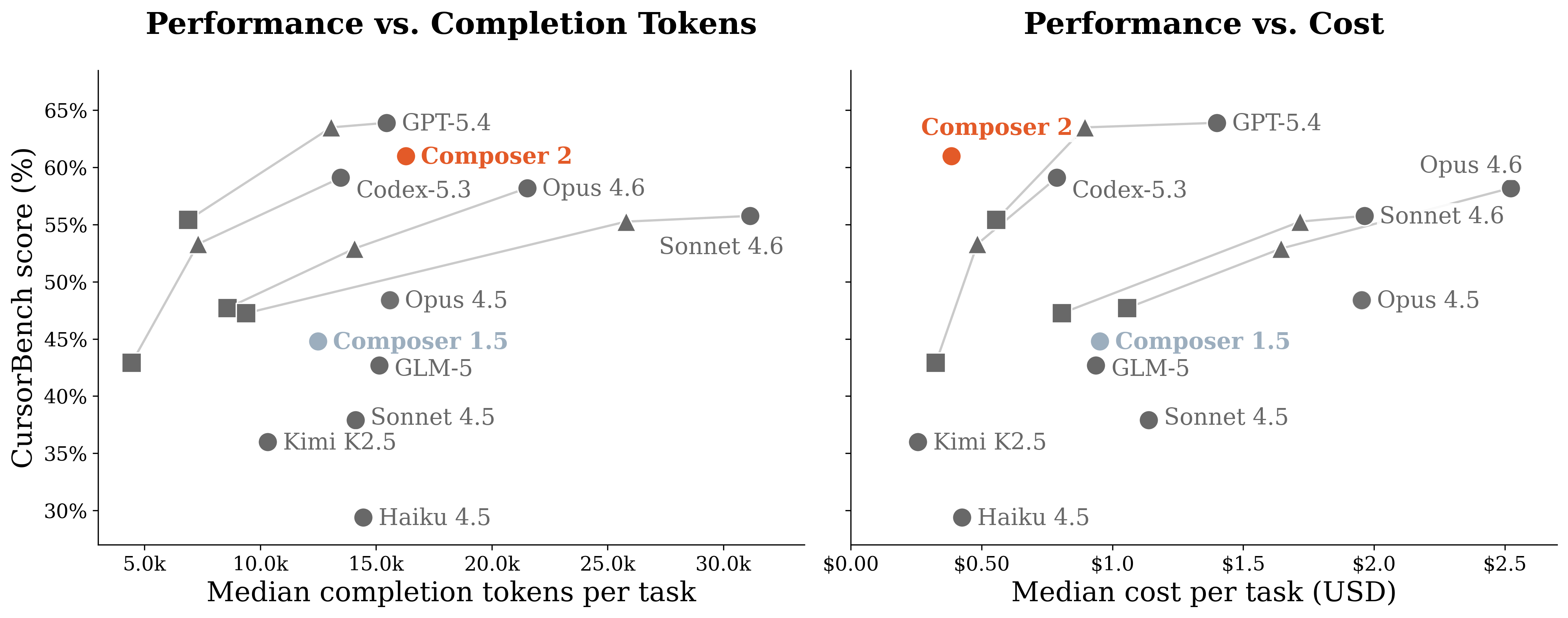}

\caption{\textbf{On CursorBench, Composer 2 achieves a superior Pareto frontier in cost while remaining highly competitive in token efficiency.} For GPT-5.4, Codex-5.3, Opus 4.6, and Sonnet 4.6, we plot the high (circle), medium (triangle), and low (square) effort variants.}
\label{fig:perf_tradeoffs}
\end{figure}

Table~\ref{tab:benchmark_results} reports the accuracy of various models on CursorBench-3.
Composer 2 achieves 61.3\%, representing a 37\% relative improvement over Composer 1.5 and a 61\% improvement over Composer 1.
Compared to its base model, Kimi K2.5, Composer 2 demonstrates a substantial accuracy boost, validating the effectiveness of our continued pretraining and reinforcement learning pipeline.
Furthermore, Composer 2 achieves accuracy competitive with the strongest frontier models despite being significantly cheaper at inference.

Figure~\ref{fig:perf_tradeoffs} contextualizes these accuracy metrics against resource consumption.
Regarding token usage, Composer 2 generates trajectories comparable in length to other models while providing frontier-level accuracy, remaining highly token-efficient relative to other frontier models operating at similar accuracy levels.

However, due to differences in active parameter counts, raw token usage does not fully capture inference efficiency.
Since we do not have access to FLOPs used by API models, we provide the median inference cost per CursorBench task in Figure~\ref{fig:perf_tradeoffs}.
Here, Composer 2 achieves a Pareto-optimal trade-off: its inference cost is similar to smaller or low-effort variants of models, while its accuracy remains competitive with much larger frontier models.
Together, these results demonstrate that domain-specialized training can yield models that are simultaneously more accurate and more cost-effective than general-purpose alternatives for the demanding requirements of real-world software engineering.

\begin{table}[t!]
    \centering
    \caption{\textbf{Benchmark results across public and internal evaluation suites.} For third-party models, we present results in an (our harness / self-reported) format where both are available. For Anthropic models on Terminal-Bench, we report the Claude Code scores from the official leaderboard in place of our harness evaluation. Overall, Composer 2 achieves accuracy competitive with the strongest frontier models.}
    \label{tab:benchmark_results}
    \begin{tabular}{l c c c}
    \toprule
    \textbf{Model} & \textbf{CursorBench} & \textbf{SWE-bench Multi.} & \textbf{Terminal-Bench} \\
    \midrule
    Composer 2 & 61.3 & 73.7 & 61.7 \\
    Composer 1.5 & 44.2 & 65.9 & 47.9 \\
    Composer 1 & 38.0 & 56.9 & 40.0 \\
    \midrule
    Opus 4.6 High & 58.2 & \makebox[2.2em][r]{75.8} / \makebox[1.8em][l]{77.8} & \makebox[2.6em][r]{58.0} / \makebox[2em][l]{65.4} \\
    Opus 4.5 High & 48.4 & \makebox[2.2em][r]{73.8} / \makebox[1.8em][l]{76.2} & \makebox[2.6em][r]{52.1} / \makebox[2em][l]{59.8} \\
    \midrule
    GPT-5.4 & 63.9 & \makebox[2.2em][r]{76.8} / \makebox[1.8em][l]{-} & \makebox[2.6em][r]{66.5\rlap{$^{\dagger}$}} / \makebox[2em][l]{75.1} \\
    GPT-5.3 Codex & 59.1 & \makebox[2.2em][r]{74.8} / \makebox[1.8em][l]{-} & \makebox[2.6em][r]{64.8\rlap{$^{\dagger}$}} / \makebox[2em][l]{77.3} \\
    GPT-5.2 & 56.5 & \makebox[2.2em][r]{68.3} / \makebox[1.8em][l]{-} & \makebox[2.6em][r]{60.5} / \makebox[2em][l]{62.2} \\
    \midrule
    GLM-5 & 42.7 & \makebox[2.2em][r]{66.9} / \makebox[1.8em][l]{73.3} & \makebox[2.6em][r]{{59.6}} / \makebox[2em][l]{56.2} \\
    Kimi K2.5 & 36.0 & \makebox[2.2em][r]{65.1} / \makebox[1.8em][l]{73.0} & \makebox[2.6em][r]{47.3} / \makebox[2em][l]{50.8} \\
    \bottomrule
    \end{tabular}
    \par\vspace{4pt}
    {\raggedright\small
    $^{\dagger}$OpenAI safety filters refused 5 GPT-5.4 and 3 GPT-5.3-Codex tasks; refused problems scored as 0.}
    \end{table}
    
\subsection{Public Benchmarks}
\label{subsec:public_benchmarks}

We further evaluate Composer 2 on two public benchmarks: SWE-bench Multilingual and Terminal-Bench (Table~\ref{tab:benchmark_results}, last two columns).
For Composer models, we compute scores using our own harness. 
For third-party models, we report results as (our harness / self-reported) where both are available; for Anthropic models on Terminal-Bench, we use the official Claude Code leaderboard scores rather than our own harness evaluations.
For SWE-bench, we simply prepend ``\textit{please solve this github issue}'' to the problem statement without instructions for writing or running test cases.
For Terminal-Bench, we augment the user prompt with solution formatting instructions on where files should be placed or environment should be set up.

On SWE-bench Multilingual, Composer 2 scores 73.7\%, a 7.8\% improvement over Composer 1.5 and 16.8\% over Composer 1.
On Terminal-Bench, Composer 2 achieves 61.7\%, improving upon Composer 1.5 by 13.8\% and Composer 1 by 21.7\%.
Against its base model, Kimi K2.5, Composer 2 achieves similar performance on SWE-bench Multilingual and considerably improved performance on Terminal-Bench.
Overall, Composer 2's performance on these public benchmarks remains highly competitive with other state-of-the-art models.
Across both benchmarks, each successive Composer version shows consistent gains, demonstrating that continued investment in both pretraining and reinforcement learning yields compounding gains for agentic software engineering.    

\vspace{-5pt}
\section{Conclusion}
\label{sec:conclusion}

Composer 2 demonstrates that strong specialized models can be trained through continued pretraining and reinforcement learning. Starting from a strong general-purpose model, a model can be specialized to achieve frontier-level performance in agentic coding. The main insight, from both an algorithmic and infrastructure point of view, is to scale training while ensuring a close domain match with the target domain. We do this through careful domain benchmarking with CursorBench, harness and environment engineering, and behavioral reward development, along with rigorous infrastructure reliability.

The results of Composer 2 are optimistic on the future improvement available through further scaling. While Composer 2 marks a steady improvement over previous versions, there are many cases where the model shows intelligence or coherence behaviors that can be clearly improved. The model trained in this work is large (1.04T parameters, 32B active) but likely smaller than other proprietary models of comparative ability. We believe there remains considerable room for development both architecturally and algorithmically.

The scope of coding agents as a tool is also expanding from interactive problems to agentic tasks that would require hours of human time \cite{measuring-ai-ability-to-complete-long-tasks}, with a general expectation that the horizon will grow quickly in the future~\cite{cursor_thirdera}. For future Composer iterations, our team is focused on expanding the ability of the model to work on these problems through training methods to handle longer problems both in the algorithms to effectively utilize longer term training signal and in the infrastructure to support faithful long-horizon problems.

\bibliography{references}
\bibliographystyle{colm2026_conference}

\appendix
\section{Contributors}
\label{app:contributors}

The Composer research team consists of: 

Aaron Chan,
Ahmed Shalaby,
Alexander Wettig,
Aman Sanger,
Andrew Zhai,
Anurag Ajay,
Ashvin Nair,
Charlie Snell,
Chen Lu,
Chen Shen,
Emily Jia,
Federico Cassano,
Hanpeng Liu,
Haoyu Chen,
Henry Wildermuth,
Jacob Jackson,
Janet Li,
Jediah Katz,
Jiajun Yao,
Joey Hejna,
Josh Warner,
Julius Vering,
Kevin Frans,
Lee Danilek,
Less Wright,
Lujing Cen,
Luke Melas-Kyriazi,
Michael Truell,
Michiel de Jong,
Naman Jain,
Nate Schmidt,
Nathan Wang,
Niklas Muennighoff,
Oleg Rybkin,
Paul Loh,
Phillip Kravtsov,
Rishabh Yadav,
Sahil Shah,
Sam Kottler,
Alexander M Rush,
Shengtong Zhang,
Shomil Jain,
Sriram Sankar,
Stefan Heule,
Stuart H. Sul,
Sualeh Asif,
Victor Rong,
Wanqi Zhu,
William Lin,
Yuchen Wu,
Yuri Volkov,
Yury Zemlyanskiy,
Zack Holbrook,
Zhiyuan Zhang

\section{Base Model Selection}
\label{sec:basemodelselection}

Before training, we evaluated several potential open-source base models including GLM-5~\cite{glm5}, Kimi K2.5~\cite{kimik2_5}, and DeepSeek V3.2~\cite{deepseek_v32}. Three base model evaluations contributed to our selection of Kimi K2.5:
\begin{itemize}[leftmargin=*]
    \item \textit{Coding knowledge}: We score factual knowledge with an internal benchmark called FreshBench. FreshBench is a question-answer benchmark adversarially constructed against previous Composer models. We identify turns where Composer had to read library source code or perform a web search to solve a coding task. From these traces we create question-answer pairs, validating the answers with a web searching agent.
    \item \textit{State tracking}: While editing a repository, coding agents often need to understand dozens of past file edits before taking an action.     
     LoCoDiff~\cite{locodiff} is a benchmark that asks the model to recreate the state of a file after many diffs, an important base skill for model long-term memory. State tracking is an internal benchmark similar to LoCoDiff built from our monorepo.
     Instead of measuring raw accuracy, which we found sensitive to single-character errors, we report the average character-level distance.
    \item \textit{Codebase perplexity}: We measure perplexity to determine the coding intelligence of the base model.
    We use our private monorepo as an uncontaminated source, concatenating the files alphabetically and computing the sum of the negative log-likelihoods over a rolling window.

\end{itemize}

We intentionally do not consider coding agent benchmarks when testing base models. We find that such benchmarks are less predictive of final performance, as agentic and long-horizon capabilities can drastically change during the RL stage.

Table~\ref{tab:model_selection} shows the results of the analysis. All three models considered perform quite well in these experiments. We selected Kimi K2.5~\cite{kimik2_5} due to its general strong performance as well as further additional considerations such as its efficiency in our infrastructure. 

\begin{table}[t!]
\centering
\begin{tabular}{lrrr}
\toprule
Model  & FreshBench $\uparrow$ & State Tracking $\downarrow$ &  Negative Log-Likelihood $\downarrow$  \\
\midrule
DeepSeek V3.2 & 68.9\% & \textbf{66} & \textbf{11.75M} \\
\textbf{Kimi K2.5} & \textbf{83.2\%} & 86 & 13.81M \\
GLM-5 & 79.2\% & 92 & 14.11M \\
\midrule
GPT-5.4 & \textbf{92.5}\% & 103 & - \\
Claude 4.6 Opus & 88.9\% & 65 & - \\
Gemini 3 Flash & 84.5\% & \textbf{27} & - \\
Claude 4.5 Sonnet & 80.1\% & 69 & - \\
Claude 4.5 Haiku & 61.7\% & 177 & - \\
\bottomrule
\end{tabular}
\caption{\textbf{Base models evaluated on our internal benchmarks.} Negative log-likelihood is measured over our internal codebase.}
\label{tab:model_selection}
\end{table}

\section{CursorBench}
\label{app:gold_diffs}

\subsection{Streaming Prefix Detection}
\label{app:streaming_gold_diff}
The following is another example CursorBench task.

\begin{figure}[h!]
\centering
\begin{tcolorbox}[colback=gray!5, colframe=gray!50, boxrule=0.5pt, left=6pt, right=6pt, top=4pt, bottom=4pt]
\small\textbf{Problem statement:} We're seeing a weird streaming bug in some chat responses: 
\newline
\newline
\textit{
Now I \newline
Now I need to updat \newline
Now I need to update this. \newline
Now I need to update this. I ha \newline
Now I need to update this. I have the} \newline
\newline
Instead of getting proper streaming deltas, we get repeated growing prefixes like the snippet. I think this happens mostly inside think tokens. I want to know how common this is. Look at 954 response json files in @logs folder
\end{tcolorbox}
\vspace{-0.5em}
\caption{\textbf{Example CursorBench task.} The agent must infer the failure mode from a partial symptom report, write a heuristic detection algorithm over 954 heterogeneous chat responses, and carefully tune that heuristic to recover an exact count of malformed prefix-streaming cases without overcounting normal incremental output. Additionally, a variant of the bug produces an ``interleave stutter'' where the initial prefix chain is only two lines long before stabilizing into a repeating line with incrementing repetitions and agent must carefully examine chat responses to discover this.}
\label{fig:cursorbench_example_streaming}
\end{figure}

The following listing shows the algorithmic core of the reference diff for this task.

\begin{lstlisting}[language=Python, basicstyle=\ttfamily\scriptsize]
MIN_CHAIN = 3
MIN_SEED_LEN = 2
MAX_SEED_LEN = 50

def find_prefix_chain(text: str) -> tuple[int, str] | None:
    if len(text) < 10:
        return None
    first_nl = text.find("\n")
    if first_nl < MIN_SEED_LEN or first_nl > MAX_SEED_LEN:
        return None
    seed = text[:first_nl]
    needle = "\n" + seed
    starts = [0]
    pos = 0
    while True:
        idx = text.find(needle, pos)
        if idx == -1:
            break
        starts.append(idx + 1)
        pos = idx + 1
    if len(starts) < MIN_CHAIN:
        return None
    ends = [s - 1 for s in starts[1:]] + [len(text)]
    chunks = [text[s:e] for s, e in zip(starts, ends)]
    chain = 1
    for i in range(len(chunks) - 1):
        cur, nxt = chunks[i], chunks[i + 1]
        if len(cur) < len(nxt) and nxt.startswith(cur):
            chain += 1
        else:
            break
    return (chain, seed) if chain >= MIN_CHAIN else None

def iter_think_blocks(text: str):
    pos = 0
    while True:
        open_idx = text.find("<think>", pos)
        if open_idx == -1:
            return
        close_idx = text.find("</think>", open_idx)
        if close_idx == -1:
            yield text[open_idx + 7:].lstrip("\n")
            return
        yield text[open_idx + 7:close_idx].lstrip("\n")
        pos = close_idx + 8

def has_prefix_streaming_bug(chat_response: str) -> bool:
    return any(
        find_prefix_chain(block) is not None
        for block in iter_think_blocks(chat_response)
    )
\end{lstlisting}

\end{document}